# Raman Spectroscopy, Photocatalytic Degradation and Stabilization of Atomically Thin Chromium Triiodide

Dmitry Shcherbakov[1], Petr Stepanov[1], Daniel Weber[2], Yaxian Wang[3], Jin Hu[4], Yanglin Zhu[4], Kenji Watanabe[5], Takashi Taniguchi[5], Zhiqiang Mao[4], Wolfgang Windl[3], Joshua Goldberger[2], Marc Bockrath[1], Chun Ning Lau[1]*

[1] Department of Physics, The Ohio State University, Columbus, OH 43210
[2] Department of Chemistry and Biochemistry, The Ohio State University, Columbus, OH 43210
[3] Department of Materials Science and Engineering, The Ohio State University, OH 43210
[4] Department of Physics and Engineering Physics, Tulane University, New Orleans, LA 70118
[5] National Institute for Materials Science, 1-1 Namiki Tsukuba Ibaraki 305-0044 Japan.

**Abstract**

As a 2D ferromagnetic semiconductor with magnetic ordering, atomically thin chromium triiodide is the latest addition to the family of two-dimensional (2D) materials. However, realistic exploration of $CrI_3$-based devices and heterostructures is challenging, due to its extreme instability under ambient conditions. Here we present Raman characterization of $CrI_3$, and demonstrate that the main degradation pathway of $CrI_3$ is the photocatalytic substitution of iodine by water. While simple encapsulation by $Al_2O_3$, PMMA and hexagonal BN (hBN) only leads to modest reduction in degradation rate, minimizing exposure of light markedly improves stability, and $CrI_3$ sheets sandwiched between hBN layers are air-stable for >10 days. By monitoring the transfer characteristics of $CrI_3$/graphene heterostructure over the course of degradation, we show that the aquachromium solution hole-dopes graphene.

TOC Graphic:

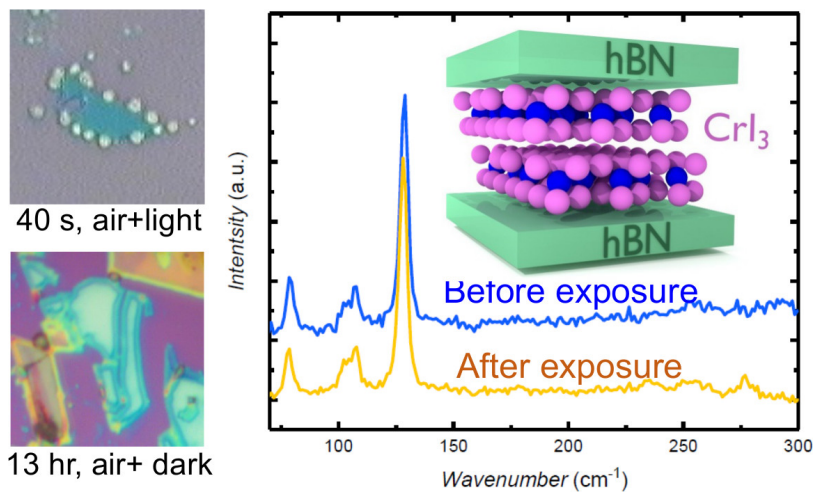

* Email: lau.232@osu.edu

The advent of two-dimensional (2D) materials [1] has opened the door for almost limitless exploration of materials with unique functionalities. 2D counterparts of electronic phases that are present in the bulk have been discovered, such as Ising superconductors [2,3], charge density waves [4,5], and topological semimetals [6]. Moreover, by forming van der Waals stacks of atomic layers of different materials, the heterostructures' properties can be tailored by quantum confinement and proximity effects.

Very recently, magnetic orders in monolayer and bilayer chromium triiodide $CrI_3$ have been reported [7], thus adding magnetism to the collection of 2D counterparts of non-Fermi liquid electronic phases. Bulk $CrI_3$ is a ferromagnetic semiconductor, with out-of-plane magnetic moments, a band gap of 1.2 eV and a Curie temperature $T_C$ =61 K. It also is reported to undergo a structural phase transition ~ 210-220 K, switching from a monoclinic lattice at high temperature to rhombohedral at low temperature [8]. The ferromagnetism survives in monolayers, with a band gap that increases to 1.5 eV [7,8], and $T_C$ that decreases to 45 K. In bilayer, $CrI_3$ was found to be antiferromagnetic [9,10]. $CrI_3$-based heterostructures were shown to control spin and valley pseudospin in $WSe_2$ [11], and give rise to giant tunneling resistance [12,13]. These cutting-edge studies underscore $CrI_3$'s promise for fundamental exploration of electron correlation in low dimensions and interfaces, as well as spintronics and magnetoelectronics applications.

Despite the intense efforts, a significant obstacle in realizing the promise of $CrI_3$ is its instability -- under ambient conditions, it turns into a liquid within minutes, thus severely hampering fabrication and exploration of any $CrI_3$-based devices. In this Letter, we present the first characterization of $CrI_3$ by polarized Raman spectroscopy, and a systematic study of the chemical reactivity of $CrI_3$ under various conditions, particularly addressing chemical stability under standard fabrication conditions for making electrical contacts, such as resist baking and exposure to light. We establish that much of the degradation of atomically thin $CrI_3$ arises from photo-induced ligand substitution with $H_2O$, leading to the formation of aquachromium iodides that are highly hygroscopic and water soluble. By experimenting with different encapsulation materials such as polymethyl methacrylate (PMMA), $Al_2O_3$ and hBN, we demonstrate that $CrI_3$ stability is markedly improved if its exposure to light is minimized. Best stability, >10 days under ambient conditions and >15 hours under focused light, is achieved by sandwiching few-layer $CrI_3$ sheets within large hBN flakes. As a demonstration, we fabricate hBN/bilayer graphene/$CrI_3$ heterostructures. As $CrI_3$ undergoes photo-induced aquation, graphene becomes highly hole-doped with reduced mobility, suggesting that doping by aquachromium solution can be used to tune the charge density of graphene.

$CrI_3$ bulk crystals are synthesized via chemical vapor transport using elemental Cr and I in a sealed tube [8]. The as-grown $CrI_3$ samples crystallize into a $C2/m$ monoclinic lattice having a single $CrI_3$ layer per unit cell (Figure 1a) based on X-ray diffraction (Figure S1).[14] These bulk $CrI_3$ crystals exhibit a ferromagnetic transition temperature at 61 K (Figure S2),[14] which is consistent with the literature[8,15]. A previous study on single $CrI_3$ crystals have reported that the room temperature crystal structure has a $C2/m$ space group, which undergoes a structural phase transition below 210-220 K to transform into the $R$-3 space group [8]. These studies show that this structural phase transition is accompanied by a transition in the magnetic susceptibility ($d\chi/dT$). In our samples, a similar transition is observed in the magnetic susceptibility data (Figure S2).[14]

Next, we also characterized the polarized Raman spectra (Figure 1b) of these crystals. The unpolarized Raman spectrum of bulk $CrI_3$ crystals displays a series of peaks at frequencies of 77.1, 100-110, 128.4, 231.7, and a weak, broad excitation at 253.2 cm$^{-1}$. The peaks between 100 to 110 cm$^{-1}$ were best fit to four peaks that had frequencies of 101.1, 103.5, 106 and 108 cm$^{-}$

[1], respectively (Figure S3). In the *C*2/*m* space group ($C_{2h}$ point group) there should be 24 total vibrational branches at the Γ point in the Brillouin zone, of which only 12 are Raman active (6 $A_g$ + 6 $B_g$). We collected polarized (0° and 90°) backscattered Raman measurements oriented along the out of plane direction to elucidate the symmetry of each Raman mode (Figure 1b). Since the out of plane direction of the crystal is tilted from the *c*-axis, the cross-polarized and co-polarized intensities were compared to those predicted from first principles calculations (Figure 1c). Within our accessible experimental spectral range (>60 cm$^{-1}$), these calculations predict $A_g$ modes to appear at 85.8, 103.6, 109.9, 132.8, and 231.7 cm$^{-1}$, and $B_g$ modes to appear at 95.8, 105.2, 107.5, 214.1 and 232.6 cm$^{-1}$. These predicted modes are very close, within 10 cm$^{-1}$, to our observed spectra. In general, the *$A_g$* modes are predicted to disappear in intensity under cross-polarization, whereas the *$B_g$* modes remain. Indeed, in our experimental spectra, we observe almost complete disappearance of the 77.1 cm$^{-1}$ peak, which indicates that it is an *$A_g$* mode. In addition there is a significant decrease in the intensity of the Raman mode at 126.7 cm$^{-1}$. The observed residual intensity is likely due to slight deviations from 90° in the polarizer angles, and not the presence of a *$B_g$* mode as no other Raman mode is predicted within 25 cm$^{-1}$. Indeed, first principles simulation of the Raman spectra show residual intensity appearing with a 10° deviation from cross-polarization (Figure 1c). Finally, the experimentally observed weak polarization dependence in modes between 101 to 108 cm$^{-1}$ and at 231.7 cm$^{-1}$ suggests that they are of predominantly *$B_g$* origin. We attribute the 253.2 cm$^{-1}$ mode to be a two-phonon Raman mode of the extremely intense 126.7 cm$^{-1}$ vibration, on account of the fact that it occurs at a wavenumber at twice its energy, its broadness, and because similar two-phonon Raman modes have been assigned in $CrCl_3$ and $CrBr_3$ [16].

The bulk $CrI_3$ crystals are exfoliated onto $Si/SiO_2$ substrates in an Ar-filled glove box having $O_2$ and $H_2O$ concentrations <0.1 ppm. $CrI_3$/hBN heterostructures are fabricated by a dry transfer technique [17, 18]. Unless indicated otherwise, all optical microscope images are taken inside the glove box. Figure 1d displays optical microscope of a typical $CrI_3$ flake, and its thickness measured from atomic force microscope (AFM) measurements. Atomic steps ~ 0.7 nm in height are observed, indicating monolayer step size.

To study its chemical reactivity, we first examine a few nm-thick $CrI_3$ flake under an optical microscope in air, and take successive images at 20-second intervals. As shown in the series of images in Figure 2a, small liquid droplets form around the flake's edges in less than 20 seconds. Under illumination of the microscope, these droplets grow quickly with time, and the flake all but disappears after 60 seconds. To ascertain the reactant in air that induces this degradation, we place $CrI_3$ flakes in a quartz tube with controlled gas flow. In the presence of dry $O_2$ with <100 ppm water vapor, $CrI_3$ flakes are stable for >15 hours; however, in "wet" Ar gas that is bubbled through a water flask, the flakes dissolve into droplets within tens of minutes (Figure 2b). These droplets dissolve in distilled water, leaving no visible residue. Thus we identify water as the main reactant with $CrI_3$ under ambient conditions – the $I^-$ ligand that is coordinated to $Cr^{3+}$ will be substituted by water, making the surface of the partially hydrated $CrI_{3-x}(H_2O)_x^{x+}$ 2D material, much more hygroscopic and prone to dissolution.

Surprisingly, the above substitution is observed to be photo-activated. Figure 2c displays a series of optical images of a $CrI_3$ flake that has been kept in the dark in air for 0 minutes, 5 hours, 13 hours and 48 hours, respectively. The flakes appear to be much more stable, displaying no visible alteration at ~ 5 hours. Thus light accelerates $CrI_3$'s reactivity with water by 3 orders of magnitude. In fact, by minimizing exposure to light, we are able to fabricate and handle $CrI_3$-based devices outside a glovebox.

This photo-induced reactivity of $CrI_3$ can be rationalized when considering the crystal chemistry of $CrI_3$. In the crystal structure of $CrI_3$, each $Cr^{3+}$ ion exists in octahedral coordination to the $I^-$ ligands. Complexes containing octahedrally-coordinated $Cr^{3+}$ are known to be among the most substitutionally inert species of the transition metal series, on account of the large ligand field stabilization energy of the half-filled $t_{2g}$ $d^3$ electron configuration.[19] For instance, the rate constant of water exchange of $Cr(H_2O)_6^{3+}$ is ~2 x$10^{-6}$ $M^{-1}s^{-1}$, which is the second slowest rate constant for exchange between hexaaqua (III) transition metal ions. However, it has been well established that much faster ligand substitution in $Cr^{3+}$ complexes can occur upon photoexcitation of the $Cr^{3+}$ d-d transitions.[20, 21] Clearly, the same is true for $CrI_3$. Figure S3 shows the absorption spectrum of $CrI_3$.[14] At least 4 different transitions are observed. Coincident with Tanabe-Sugano diagrams as well as previous assignments of $CrI_3$, the two lowest optical transitions that occur at 1.4 and 1.6 eV correspond to direct excitation of the $^4A_{2g}$ → $^4T_{2g}$ and $^4A_{2g}$ → $^4T_{1g}$ d-d transitions. The higher energy transitions that are observed at 2.0 and 2.7 eV correspond to ligand-to-metal charge transfer excitations. These lower energy transitions correspond to the photoexcitation of either one electron from the weakly antibonding (π*) $Cr^{3+}$ 3d $t_{2g}$ orbitals, to the strongly antibonding (σ*) $Cr^{3+}$ 3d $e_g$ set, thereby promoting ligand substitution. Further confirming that this photoactivated ligand substitution can occur through excitation of the d-d transitions, we observed that exfoliated flakes of $CrI_3$ undergo the same degradation in less than a minute when excited using a coherent 785 nm laser source.

Another common reactant in air is oxygen. At room temperature, $CrI_3$ is stable in the presence of dry $O_2$, either in the dark or in the presence of ambient light (Figure 2d). However, at 180 °C, which is the standard baking temperature for cross-linking electron beam resist, it reacts with dry $O_2$ to form $Cr_2O_3$ (Figure 2e), which appears as teal-colored thin sheets under optical microscope that are not soluble in water. This sets an upper limit temperature limit of exposure in $O_2$.

The extreme sensitivity of $CrI_3$ to light and water at room temperature, and to oxygen at annealing temperatures for photoresists presents a difficult challenge for device fabrication. Graphene encapsulation has been shown provide adequate protection[7, 13], though such devices are not suitable for transport measurements due to the metallic nature of graphene. Prior works on other air-sensitive materials have demonstrated that they can be protected by insulating encapsulating layers. For instance, black phosphorus has been shown to be air stable if encapsulated by hBN [22], $Al_2O_3$ [23] and PMMA [24, 25]. Here we investigate the effectiveness of these capping materials for protecting $CrI_3$ in ambient conditions. To this end, we deposit $Al_2O_3$ or PMMA layers, and transfer hBN sheets on top of $CrI_3$ flakes on $Si/SiO_2$ substrates; the heterostructures are then illuminated by the light of an optical microscope, or heated to 180 °C in air. As shown in Figure 3a-c, the results are rather disappointing, as the $CrI_3$ flakes under all three encapsulating materials deteriorate quickly. For PMMA and $Al_2O_3$, the degradation can be attributed to the porosity of the capping layers, demonstrating that these materials are inadequate to protect $CrI_3$ from moisture and oxygen.

The degradation of $CrI_3$ capped by hBN is surprising, as hBN is atomically flat, hosting no dangling bonds and very few defects [17], and has functioned extremely well as a protective layer for black phosphorus [22] and $NbSe_2$ [26, 27]. Here we attribute the degradation to moisture leakage through the rugged $CrI_3/SiO_2$ interface, as the $SiO_2$ substrate has ~0.5 nm corrugation [28]. To confirm this hypothesis, we sandwich a $CrI_3$ flake between large hBN sheets and test the heterostructure's stability. The $CrI_3$ flake appears to be stable under ambient conditions, with no visual degradation in 10 days. Under focused light of a halogen lamp with spectrum FWHM

~480 –750 nm and estimated intensity of ~1 W/cm$^2$, the flake is stable for >15 hours, as confirmed by Raman spectroscopy (Figure 3d). The stability of CrI$_3$ under various conditions as well as under different protective layers is summarized in Table 1.

Lastly, we examine the effect of CrI$_3$ degradation on device performance. To this end, we fabricate a CrI$_3$/bilayer graphene (BLG)/hBN heterostructure, and monitor the four-terminal conductance $G$ of BLG under illumination as a function of back gate voltage $V_{bg}$. Images of the device before and after the measurements are shown in Figure 4a-b. A source-drain bias of 0.05 mV is applied while we monitor its two-terminal conductance $G$ as a function of $V_{bg}$ (no electrochemical reaction is expected at such a small bias). The pristine device has field effect mobility ~ 2000 cm$^2$/Vs, with approximately symmetric electron and hole behaviors. The Dirac point is located at $V_{bg}$ ~ 10 V (Figure 4c, red curve), indicating slight *p*-doping, which is consistent with O$_2$ adsorption, similar to that observed in carbon nanotubes [29]. Upon illumination in air, within the first 5 minutes, graphene's Dirac point shifts progressively to negative voltages, indicating electron doping. Interestingly, its hole mobility stays approximately constant, while the electron mobility steadily decreases. After 10 minutes of illumination, the Dirac point starts to shift to the right, suggesting large hole doping, while both electron and hole mobility deteriorate. In ~ 20 minutes, the DP has shifted to > 60 V, and hole and electron mobility degrades by an order of magnitude to ~280 and 200 cm$^2$/Vs, respectively (Figure 4c). We attribute the final hole doping of the device to the formation of aquachromium solution and the dissociation of the ions (*e.g.* [Cr(H$_2$O)$_6$]$^{3+}$ and I$^-$). In other words, the solution becomes a salt solution with dissociated large cations and small anions that form an electric double layer. The I$^-$ ions accumulate on the graphene surface; the proximity of the negatively charged ions drives electron from graphene, thus highly hole-dopes graphene while providing significant additional scattering, leading to reducing charge carrier mobility.

In summary, we demonstrate that the main degradation pathway of CrI$_3$ is the photocatalytic formation of aquachromium solutions under ambient conditions, and thermocatalytic conversion to Cr$_2$O$_3$ in O$_2$. Such degradation can be prevented by sandwiching atomically thin CrI$_3$ between hBN sheets. Interestingly, aquachromium solutions can form an electric double layer, similar to that in an ionic liquid,[30] that tunes the charge density of 2D materials.

Support Information Available
Refinement of parameters for XRD diffraction data, UV-vis spectrum and temperature dependence of magnetic susceptibility of bulk CrI$_3$ crystals, fitting of Raman peaks, and details of DFT calculation.

|  | Air | Argon | $O_2$ | Humid Argon |
|---|---|---|---|---|
| Dark | 5 hours | Stable | > 15 hours | 15 min |
| Light | 5 sec | Stable | > 15 hours | < 5 min |
| Heat ($180^0$ C) | < 15 min | Stable | < 15 min | < 15 min |

|  | $Al_2O_3$ | PMMA | hBN superstrate | hBN sandwich |
|---|---|---|---|---|
| *Air + focused light* | 20 sec | 20 sec | 10 min | >15 hours (>10 days under ambient condition) |

Table 1. Stability of $CrI_3$ under various conditions and protective layers. The time corresponds to the first visible signs of degradation observed under optical microscope.

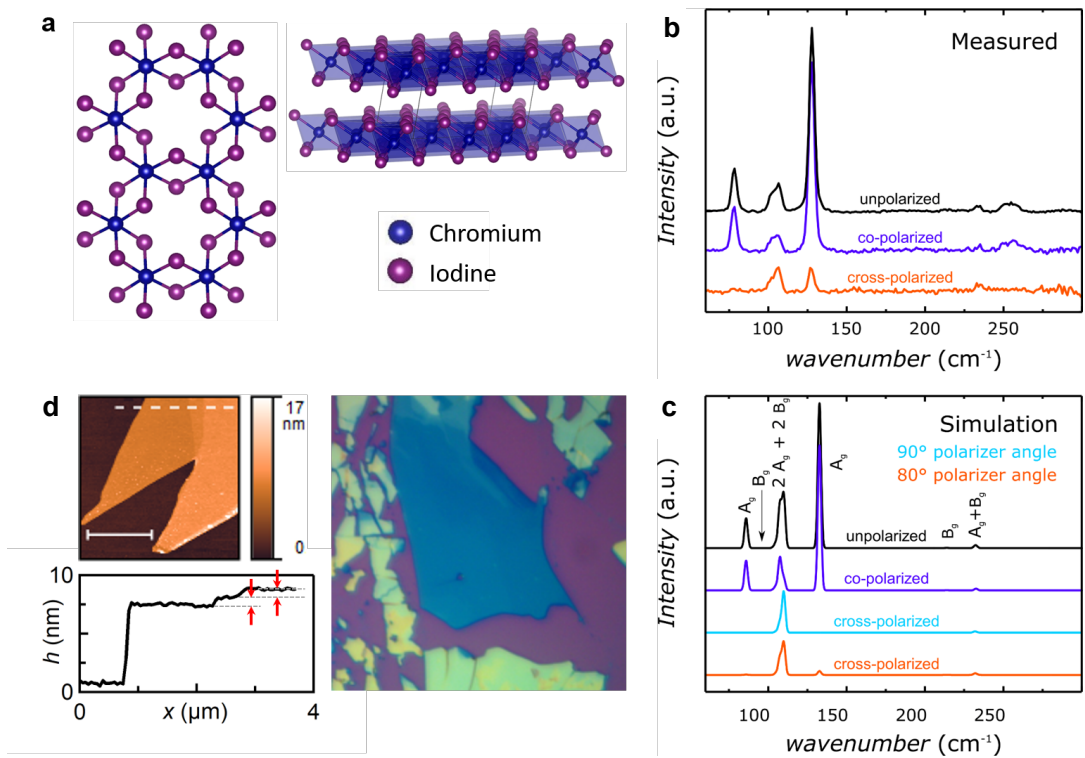

Figure 1. Crystal structure and characterization of $CrI_3$ at room temperature. (a) Schematic *C2/m* crystal structure of $CrI_3$. Left: top view of a single layer. Right: side view of bilayer. (b) Measured polarization-dependent Raman spectra of $CrI_3$. (c) DFT calculation of Raman spectra. (d). (Left panels) AFM topography of a few-layer $CrI_3$ flake taken in inert atmosphere, and trace along the dotted line in the top panel. Scale bar: 2 μm. The dotted lines and red arrows indicate step heights of 0.7 nm. (Right Panel) Optical image of a few-layer $CrI_3$ flake.

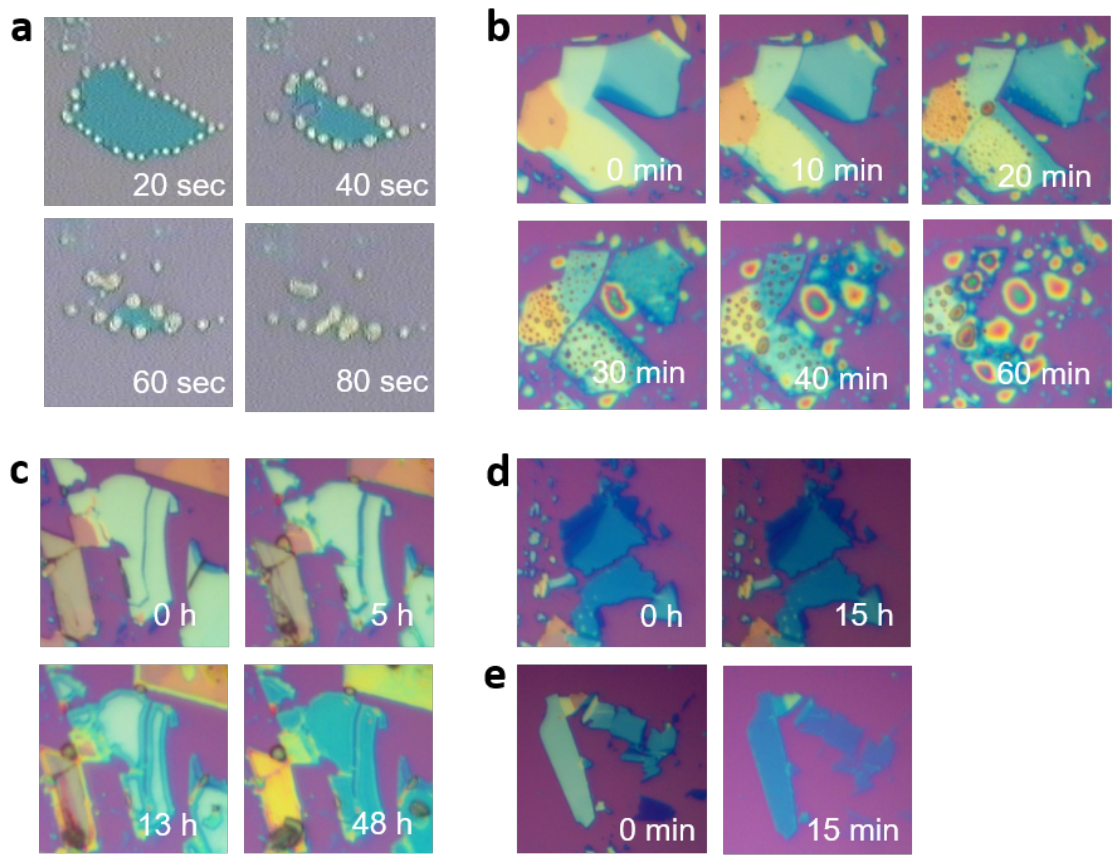

Figure 2. Micrographs of few-layer CrI3 flakes at different stages of degradation under different conditions. (a) A flake in air illuminated with light focused by a 60x microscope lens. (b) Degradation in humid argon (c) Degradation in air in the dark (d) Stability in oxygen in ambient light at room temperature. (e) Degradation in Oxygen at 180 °C.

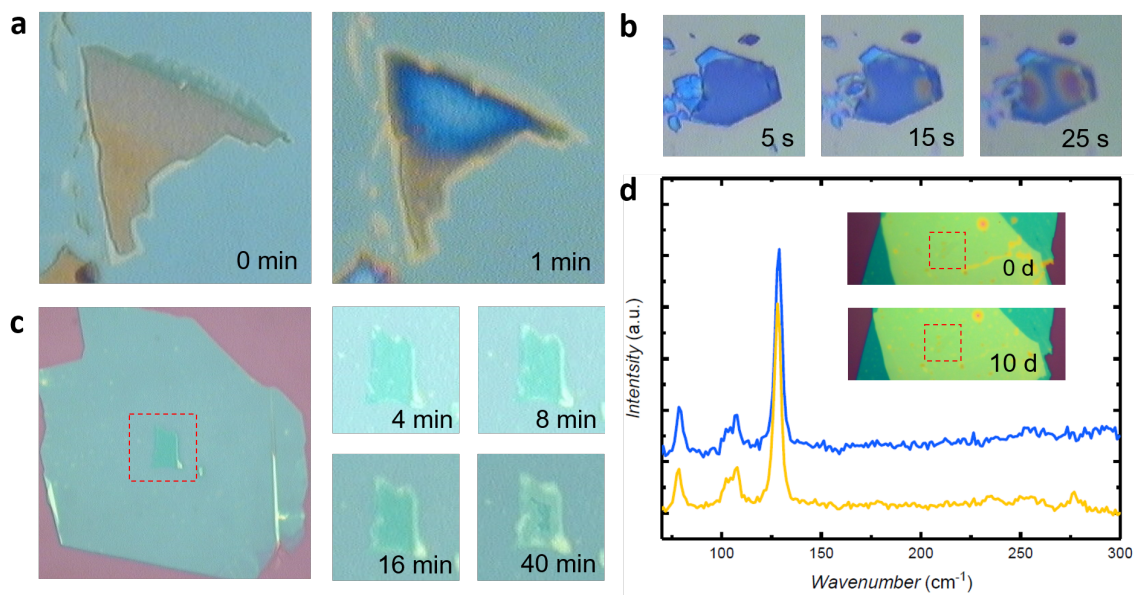

Figure 3. Stability of CrI3 flakes under different capping materials. (a) Degradation under 50 nm of evaporated $Al_2O_3$ (b) Degradation under 300 nm of PMMA (c) $CrI_3$ on $SiO_2$ substrate and covered with 20 nm-thick hBN. The panels correspond to, in clockwise: direction, 4, 8, 16 and 40 min of exposure to focused light in air, respectively. (d) Raman spectra of few-layer $CrI_3$ flake encapsulated between two hBN sheets that are ~20 and ~30 nm thick, respectively. Data acquired before (blue) and after (yellow) 10 days of exposure to ambient conditions in the dark and 15 hours under focused light (traces are offset for clarity). Insets: optical images of the flake before and after exposure.

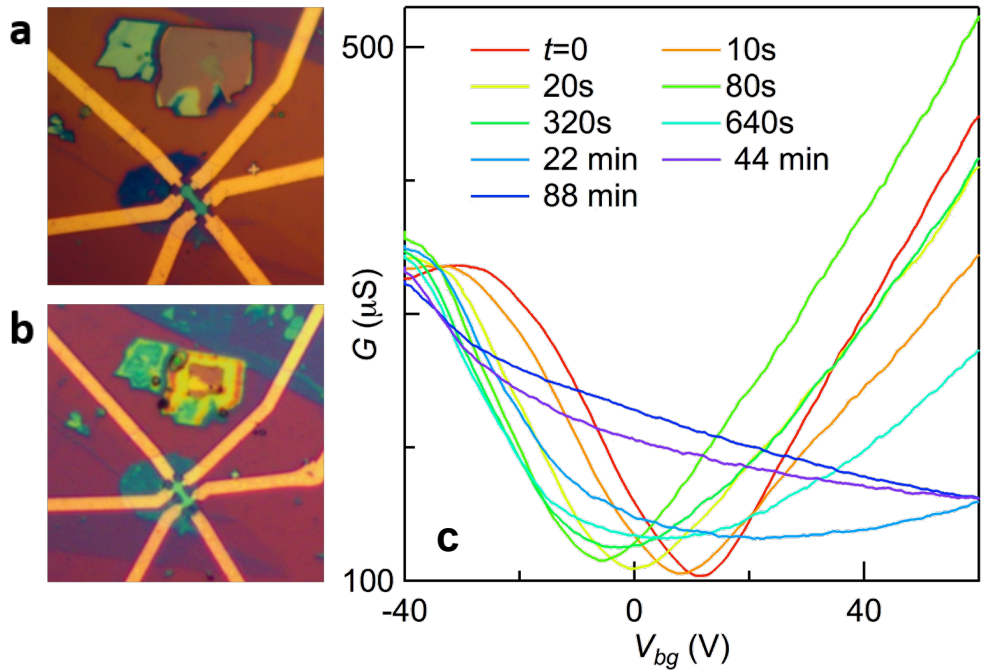

Figure 4. Optical image and transfer characteristics of a bilayer graphene sheet sandwiched between hBN and $CrI_3$. (a-b) Optical micrographs of the pristine device and after measurements. The neighboring flakes have degraded, and the device color changed, thus demonstrating the impact of air under illumination. (c) Device conductance $G$ vs back gate voltage $V_{bg}$ over the course of degradation by exposure to light in air.